\documentclass[acmsmall]{acmart}
\usepackage{amsmath}

\usepackage[utf8]{inputenc}
\usepackage[T1]{fontenc}
\usepackage{amssymb}
\usepackage{graphicx}
\usepackage{wrapfig}
\usepackage{fontawesome}
\usepackage{booktabs}
\usepackage{textcomp}
\usepackage{listings}
\usepackage[noend]{algpseudocode}
\usepackage{array}
\usepackage{multirow}
\usepackage{pifont}
\usepackage{afterpage}
\usepackage{xspace}
\usepackage{enumitem}
\usepackage[most]{tcolorbox}

\usepackage{cleveref}
\usepackage{url}
\usepackage{tabularx}
\usepackage[table]{xcolor}
\usepackage{hyperref}
\definecolor{lb}{HTML}{D8F3FD} 

\AtBeginDocument{%
  }

\definecolor{lightgray}{rgb}{0.95, 0.95, 0.95}

\newcommand{\tool}{\textsc{Ace}\xspace}
\newcommand{\benchmark}{\textsc{CEBench}\xspace}
\newcommand{\benchmarkV}{\textsc{CEBench-verified}\xspace}

\setcopyright{cc}
\setcctype{by}
\acmDOI{10.1145/3832234}
\acmYear{2026}
\acmJournal{PACMSE}
\acmVolume{3}
\acmNumber{ISSTA}
\acmArticle{ISSTA143}  
\acmMonth{10}
\acmSubmissionID{issta26main-p1446-p}
\received{2026-01-30}
\received[accepted]{2026-06-25}

\begin{document}

\title{No One-Size-Fits-All: Adaptive Code Editing with Feature-Based Strategy Selection}

\author{Jun Wan}
\email{22451014@zju.edu.cn}
\orcid{0009-0006-3294-688X}
\affiliation{%
  \department {School of Software Technology}
  \institution{Zhejiang University}
  \city{Hangzhou}
  \state{Zhejiang}
  \country{China}
}
\author{Zhongxin Liu}
\orcid{0000-0002-1981-1626}
\email{liu\_zx@zju.edu.cn}
\affiliation{%
  \department{College of Computer Science and Technology and The State Key Laboratory of Blockchain and Data Security}
  \institution{Zhejiang University}
  \city{Hangzhou}
  \state{Zhejiang}
  \country{China}
}

\author{Dajun Chen}
\email{chendajun.cdj@antgroup.co}
\orcid{0009-0000-9532-7636}
\affiliation{%
  \institution{Ant Group}
  \city{Hangzhou}
  \state{Zhejiang}
  \country{China}
}

\author{Wei Jiang}
\email{jonny.jw@antgroup.co}
\orcid{0009-0003-6605-9793}
\affiliation{%
  \institution{Ant Group}
  \city{Hangzhou}
  \state{Zhejiang}
  \country{China}
}

\author{Yong Li}
\email{liyong.liy@antgroup.co}
\orcid{0009-0005-1664-6425}
\affiliation{%
  \institution{Ant Group}
  \city{Hangzhou}
  \state{Zhejiang}
  \country{China}
}

\author{Xiaoxue Ren}
\orcid{0000-0002-5526-1617}
\authornote{Corresponding author.}
\email{xxren@zju.edu.cn}
\affiliation{%
\department {School of Software Technology}
  \institution{Zhejiang University}
  \city{Hangzhou}
  \state{Zhejiang}
  \country{China}
}

\begin{abstract}

Large language models have shown strong potential in automating code editing (CE) tasks, yet most existing systems rely on a fixed CE strategy (i.e., one-step, search-replace, and plan-apply) across diverse scenarios. In practice, CE tasks vary widely in code context, edit locality, and semantic complexity, causing no single strategy to perform consistently well. 
Moreover, systematic studies on the relationship between strategies and CE tasks remain scarce, limiting further optimization of LLM-based code editing strategies.
To address this gap, we construct \benchmark, comprising 500 real-world CE tasks curated from SWE-bench. Then we compare three representative CE strategies (one-step, search-replace, plan-apply). 
Our study reveals two key findings: (1) each strategy exhibits distinct advantages for specific CE tasks, with no universal optimality; (2) strategy effectiveness is significantly influenced by four core task features: \emph{total lines of code}, \emph{modification type}, \emph{modification scope}, and \emph{task type}.
Motivated by these findings, we propose \tool, an adaptive code editing framework that dynamically selects optimal CE strategies based on task features. Unlike classifier-based or instruction-prompting approaches, \tool employs a fine-tuned LLM that
learns feature-strategy mappings from empirical data to recommend strategies with interpretable
rationales. 
Extensive experiments on our \benchmark-verified show that \tool outperforms state-of-the-art baselines: it boosts Pass@1 by 21.7\% while reducing token consumption by 27.3\% compared to the best baseline (AutoCodeRover). Our strategy selector achieves 72.6\% selection accuracy with 69.9\% rationale accuracy (Cohen's Kappa=0.807), providing transparent and trustworthy decision-making for developers.

\end{abstract}



\maketitle

\section{Introduction}

Code editing (CE) refers to modifying existing source code to improve its correctness, functionality, performance, or maintainability \cite{mcintosh2016empirical, ram2018makes, Liang2025fuseapplybench}. Unlike code generation from scratch, CE operates on pre-existing codebases and typically involves localized modifications such as bug fixing, refactoring, and feature extension. As modern software systems undergo continuous evolution, CE has become a fundamental activity throughout the software development lifecycle \cite{mcintosh2016empirical, ram2018makes, chakraborty2020codit}.
Effective CE is increasingly critical as software systems grow in scale and complexity. Large codebases maintained over extended periods require frequent updates to address bugs, security vulnerabilities, and evolving requirements. The quality of CE directly influences software reliability, developer productivity, and long-term maintainability \cite{mcintosh2016empirical, ram2018makes, chakraborty2020codit}. However, CE remains challenging and time-consuming. Successfully editing code requires a deep understanding of program semantics, design assumptions, and module dependencies. Even small changes may introduce unintended side effects in interconnected codebases, substantially increasing cognitive workload. Studies show developers spend over six hours per week on CE tasks \cite{li2023codeeditor}, underscoring both its prevalence and importance.


Large language models (LLMs) have achieved significant advances in automating software engineering tasks
~\cite{nam2024using,bouzenia2024repairagent,yang2024swe,chew2023llm}. Prior work has demonstrated LLM capabilities across diverse activities, including code generation~\cite{Peng2025coffe}, program comprehension~\cite{comprehensive2024gao}, bug fixing~\cite{jimenez2024swebench}, and test case generation~\cite{identify2024xing}, all typically guided by natural language instructions. Notably, many of these tasks fundamentally involve CE rather than generating programs from scratch, as developers iteratively modify existing codebases throughout the development lifecycle. This observation has motivated recent research on LLM-based CE techniques that apply targeted modifications to existing code based on natural language instructions~\cite{Shrivastava2023Repository}.


\begin{wrapfigure}{r}{0.5\textwidth} 
  \centering
  \vspace{-10pt} 
  \includegraphics[width=0.5\textwidth]{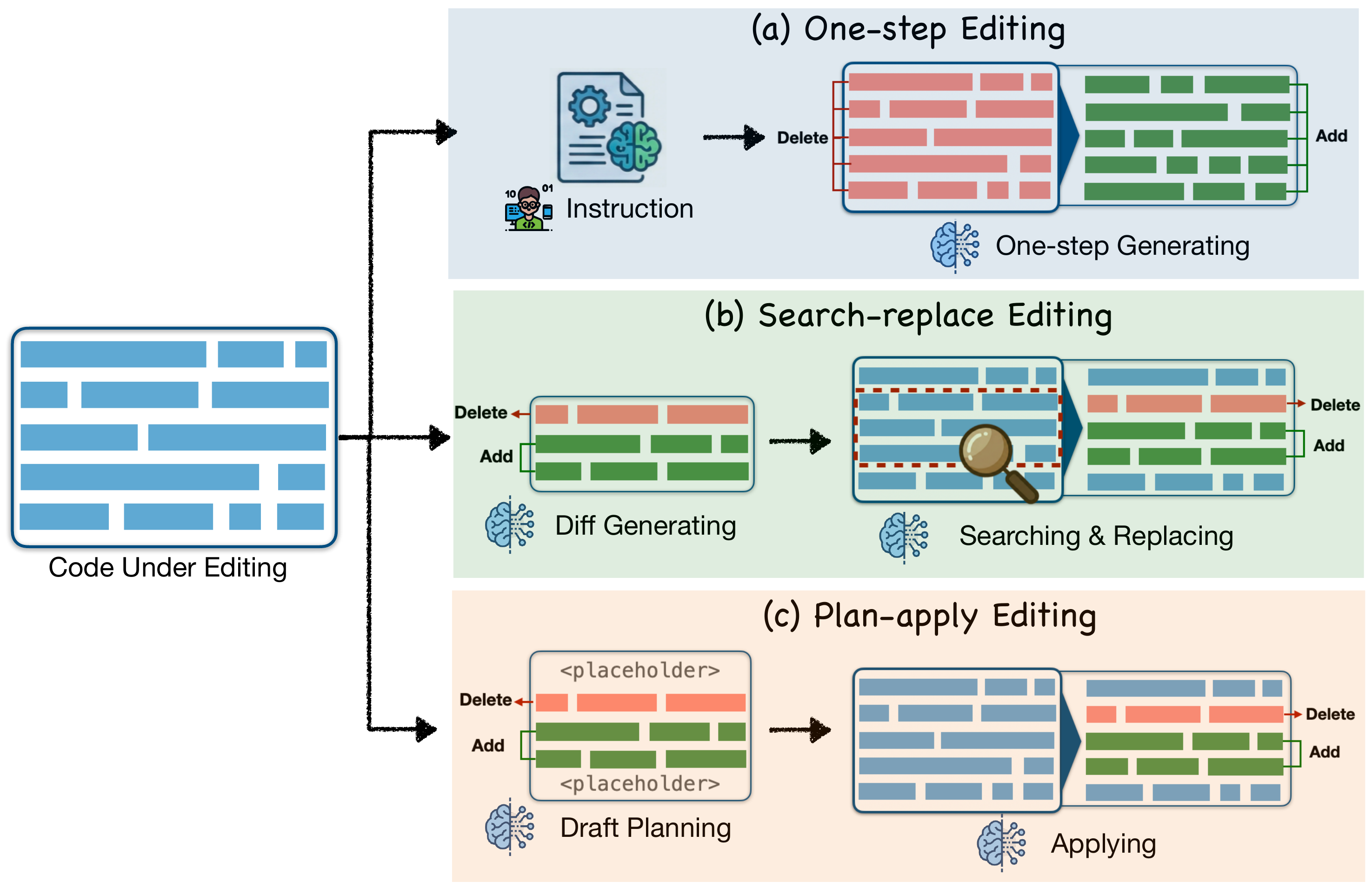}
  \caption{Modern LLM-based Code Editing Strategies}
   \label{fig:3strategies}
   \vspace{-1pt}
\end{wrapfigure}

As illustrated in Fig.~\ref{fig:3strategies}, existing CE strategies can be broadly categorized into three categories~\cite{Liang2025fuseapplybench}: one-step, search-replace, and plan-apply. 
(1) \textbf{One-step} strategy~\cite{Bairi2024codeplan} directly rewrites the code under edit using prompts that include the original code and the editing instruction (Fig.~\ref{fig:3strategies}-a). While straightforward, this strategy is highly sensitive to model capability, with weaker LLMs (e.g., <10B parameters) often producing unreliable edits~\cite{should2024qi}. Additionally, rewriting entire code sections leads to excessive token consumption and latency, particularly for large or complex codebases~\cite{Liang2025fuseapplybench}.
(2) \textbf{Search-replace} strategy~\cite{Yang2024sweagent} first generates code diffs instead of complete rewrites, and then locates and applies these diffs to the original code (Fig.~\ref{fig:3strategies}-b). While diff-based generation improves token efficiency, LLM-generated diffs often lack precision in identifying edit locations or correctly applying changes, leading to syntax or semantic errors that cause task failures~\cite{Aider2025UnifiedDiffs}.
(3) \textbf{Plan-apply} strategy~\cite{Liang2025fuseapplybench} decomposes CE tasks into planning and applying stages (Fig.~\ref{fig:3strategies}-c). By decoupling planning from execution, this strategy reduces token usage during planning and enables efficient application through optimized decoding techniques such as speculative decoding~\cite{leviathan2023fastinferencetransformersspeculative,yu2026enhancing,lu2026balancing}. It represents a recent advancement that addresses both correctness and efficiency challenges in CE, which is widely used in existing coding assistants like GitHub Copilot~\cite{Microsoft2023microsoft} and Cursor~\cite{cursor2023}.
However, it introduces additional system complexity and places higher demands on model capabilities during the applying stage. Weaker models frequently produce incomplete code (e.g., omitting necessary content or inserting placeholders), unauthorized modifications (e.g., unintended style changes), or inaccurate changes (e.g., incorrectly applying specified edits). 

Overall, the three CE strategies exhibit complementary strengths and weaknesses.
In practice, CE tasks demonstrate substantial heterogeneity in terms of codebase size, editing scope, and modification requirements. 
This heterogeneity raises fundamental questions regarding the comparative effectiveness of different strategies and the task features that determine their relative performance.
To systematically investigate the practical effectiveness of existing CE strategies, we conduct a controlled empirical study comparing one-step, search-replace, and plan-apply strategies under a unified setting.
To perform the empirical study, we construct \benchmark, a benchmark comprising 500 real-world editing tasks derived from SWE-bench~\cite{jimenez2024swebench}. 
Then, we employ three representative techniques corresponding to each strategy: CodePlan~\cite{Bairi2024codeplan} for one-step, AutoCodeRover~\cite{Zhang2024autocoderover} for search-replace, and FuseApply~\cite{Liang2025fuseapplybench} for plan-apply. Using these techniques, we conduct comparative experiments on \benchmark to address two empirical questions:
How much do strategies differ in effectiveness on CE tasks (RQ1)? And which features of CE tasks influence the relative effectiveness of these strategies (RQ2)?

Our empirical study reveals two principal findings: \textbf{(1) No single CE strategy achieves universal optimality, with each strategy exhibiting distinct advantages for different task characteristics. (2) Strategy effectiveness is task-specific and significantly influenced by four key features: \emph{total lines of code}, \emph{modification type}, \emph{modification scope}, and \emph{task type}.}


Inspired by these findings, we propose \tool, the first \underline{A}daptive \underline{C}ode \underline{E}diting framework with feature-based strategy selection to enhance both the correctness and the token efficiency of CE tasks.
Unlike classifier-based or instruction-prompting approaches, \tool employs a fine-tuned LLM that learns feature-strategy mappings from empirical data to recommend strategies with interpretable rationales.
Specifically, \tool includes three phases: feature extraction (Phase I), strategy selection (Phase II), and code editing (Phase III). First, we utilize the dataset in \benchmark to fine-tune a feature-based strategy selection model. Then, \tool employs the fine-tuned model to analyze task features and recommend the most appropriate strategy from one-step, search-replace, or plan-apply, along with interpretable rationales for the selection. 
The selected strategy then guides LLMs to perform code editing in a task-aligned manner.
By adapting strategy selection to individual task requirements rather than applying fixed approaches, \tool achieves superior performance across diverse CE scenarios while maintaining interpretability through explicit strategy justifications.

To evaluate \tool, we further construct a benchmark \benchmarkV with 650 real-world code editing tasks, including 429 single-file and 221 multi-file instances, derived from SWE-bench and SWE-bench Verified~\cite{jimenez2024swebench}.
We focus on code editing as the core repair step in end-to-end issue resolution, and our evaluation strictly aligns with this scope to ensure consistency between motivation and experimental setting.
Experimental results on \benchmarkV show that \tool significantly outperforms both fixed-strategy approaches and alternative adaptive baselines in task correctness and token efficiency. Compared to the best-performing fixed strategy (i.e., search-replace), \tool achieves 21.7\% improvement in task correctness (Pass@1) while reducing token consumption by 27.3\%. Against adaptive baselines using instruction-prompting and classifier-based selection, \tool achieves up to 75.2\% improvement in Pass@1. Our strategy selector achieves 72.6\% selection accuracy. Its rationale accuracy, as measured by manual labeling (Cohen's Kappa $= 0.807$), reaches 69.9\%. Furthermore, \tool demonstrates robust generalization across six diverse LLMs (both closed-source and open-source), consistently improving correctness by an average of 21.9\% while reducing token consumption by 22.0\%.

In summary, our contributions are as follows:
\begin{itemize}[leftmargin=*]
    \item We conduct the first empirical study that not only offers a fine-grained comparison of existing CE strategies but also provides critical insights that no single CE strategy achieves universal
optimality, and strategy effectiveness is significantly influenced by task-specific features. 
    \item We propose \tool, an adaptive code-editing framework that uses a feature-based strategy selection to enhance both correctness and token efficiency in CE tasks.
     \item We construct \benchmark and \benchmarkV from SWE-bench to compare code-editing performance. Extensive experiments compare \tool with state-of-the-art baselines to evaluate its correctness and token efficiency in code editing tasks. Additionally, further experiments validate the applicability of \tool across different LLMs for CE tasks.
    \item We open-source the replication package~\cite{ourpackage}, including the source code of \tool, dataset, and all experimental results.
\end{itemize}


\section{Background and Related Work}

\subsection{Code Editing Problem Formulation}

The CE task aims to modify the given source code $C$ according to an editing instruction $E$ to produce the desired version $C_{\text{edit}}$. 
Formally, given a set of source code snippets $\{C_1, C_2, \dots, C_n\}$ and an edit request $E$, a technique must automatically identify the locations to modify and the corresponding content changes, and then apply an editing function $f$ to generate the updated code $\{C'_1, C'_2, \dots, C'_n\}$, where each updated snippet is given by $C'_i = f(C_i, E)$. 
In traditional approaches, the editing function $f$ typically relies on manually defined rules or static analysis tools ~\cite{Zhu2021a}. 
In LLM-based CE techniques, the LLM can serve as the editing function $f$ to automatically generate the required modifications and thereby achieve higher levels of automation and flexibility.

\subsection{Code Editing Technique}
Code editing has evolved from traditional rule-based/statistical methods to modern LLM-driven approaches, with core advancements focusing on flexibility, generalization, and context awareness. We structure the related work into non-LLM-based methods and LLM-based methods to eliminate redundancy and clarify the technical lineage.

\subsubsection{Non-LLM-Based Code Editing}

Early CE techniques primarily depend on rule-based or statistical approaches. For example, template-driven repair techniques could handle common syntax errors or simple statement-level modifications \cite{Chakraborty2022codit}, \cite{Zhu2021a}. While effective in constrained settings, these methods lacked flexibility and generalization, as they were limited by the coverage of manually defined rules. The emergence of pretraining and transformer-based architectures brought about a significant shift in the field. For instance, \cite{Jiang2021cure} demonstrated that pretrained models can significantly outperform traditional methods in generating valid patches. Subsequent studies, such as \cite{Wang2021codet5} \cite{Zhang2023coditt5}, extended this line of work. 


\subsubsection{LLM-Based Code Editing}
Current LLM-based CE techniques can be classified into three main types of strategies:

\noindent\textbf{One-step Editing. \cite{Bairi2024codeplan}  \cite{li2023codeeditor} \cite{pian2025you}:} The one-step strategy generates fully modified code in a single inference pass. CodePlan~\cite{Bairi2024codeplan} targets repository-level tasks such as feature additions and cross-module refactorings, integrating repository context and dependency analysis to enhance edit coherence. 
GRACE~\cite{Gupta2023grace} fine-tunes LLMs using prompts containing related code updates to improve edit quality.
CodeEditor~\cite{li2023codeeditor} caters to repetitive editing like refactoring, with a specialized pre-training task that transforms mutated snippets into real-world ground truth.
JLED \cite{pian2025you} further emphasizes the joint modeling of edit localization and code modification, highlighting the importance of context-aware and instruction-following capabilities in practical CE scenarios.


\noindent\textbf{Search-replace Editing~\cite{Wang2025openhands} \cite{Yang2024sweagent} \cite{Zhang2024autocoderover}:} A two-step design: first locating target code snippets in the original source and then substituting them with modified diffs. AutoCodeRover~\cite{Zhang2024autocoderover} serves as a representative method tailored to localized modification tasks such as bug fixes and snippet optimization. It leverages program structure search and fault localization to enhance the accuracy of target snippet identification. 
CoEdPilot~\cite{liu2024coedpilot} leverages LLMs to identify relevant prior edits and predict change locations, enabling more precise and context-aware editing.
PEACE~\cite{ren2025peace} first constructs a dependency-aware optimizing function sequence via call graph analysis and relevance filtering, then identifies valid associated edits through semantic assessment of historical commits, and iteratively generates optimized snippets for target functions. OpenHands~\cite{Wang2025openhands} extends this paradigm by integrating multi-turn interaction, enabling more flexible snippet refinement for complex local edits. Compared to the one-step strategy, it reduces token usage and mitigates context-related issues. However, it is highly sensitive to search-stage errors, such as misinterpreting locations, and often overlooks inter-code dependencies due to isolated snippet generation.

\noindent\textbf{Plan-apply Editing \cite{Kortix2024fast} \cite{Liang2025fuseapplybench}:}  A more recent two-step paradigm, adopted by systems such as Cursor \cite{cursor2023} and Microsoft Copilot \cite{Microsoft2023microsoft}. 
The plan-apply strategy splits editing into two stages: a large LLM generates a modification draft with placeholders, and a smaller LLM fills in placeholders with original code. FuseApply~\cite{Liang2025fuseapplybench} is a benchmarking method for this paradigm, designed for large-scale CE tasks and explicitly addressing code-edit apply bottlenecks, e.g., unauthorized changes. Fast Apply~\cite{Kortix2024fast} optimizes the applying stage with fine-tuned small models, enhancing efficiency for multi-function modification tasks. This design improves edit precision by separating logic planning from code application, but introduces pipeline complexity, intermediate token overhead, and risks of placeholder misinterpretation.

All three strategies share a critical limitation: they adopt a fixed architecture for all tasks, ignoring the diverse requirements of different scenarios, resulting in inefficiency or reduced accuracy. To address this gap, we propose \tool, which dynamically selects optimal strategies based on task and source code, addressing the mismatch between fixed strategies and real-world editing requirements.


\section{Empirical Study Setup}

\subsection{Research Questions}
\label{sec: empirical Research question}
Our empirical study is structured around the following research questions:

\textbf{RQ1: How much do strategies differ in task correctness and token efficiency on code editing?}
This research question aims to quantify differences in effectiveness among the three CE strategies (i.e., one-step, search-replace, and plan-apply) in terms of task correctness and output token consumption, providing a clear empirical understanding of their relative performance.
Specifically, we adopt the evaluation methods described in Sec.~\ref{sec:empirical baseline}. 
Since Claude-Opus-4.1 exhibits strong performance on SWE-bench, all methods are executed using the same LLM Claude-Opus-4.1~\cite{Anthropic2025ClaudeOpus4} on our \benchmark (see Sec.~\ref{sec: empirical data collection}). 
For evaluation, we use pass@1 to measure editing success rate and output tokens to measure token efficiency. Additionally, we define a new metric, IsGT (see Sec.~\ref{sec:empirical metirc}), to identify which strategy should be treated as the ground-truth (GT) reference for each task among all CE strategies.

\textbf{RQ2: Which features of code editing tasks influence the relative effectiveness of these strategies?}
This research question seeks to identify the highly correlated task features that affect the performance of different editing strategies, thereby providing insights into what and why certain strategies outperform others.
To address RQ2, we extract task features from successfully solved cases using an LLM, then perform a correlation analysis (Sec.~\ref{sec:empirical feature analysis}) to identify features that significantly affect strategy performance.


\subsection{Method of Empirical Study}
\subsubsection{Techniques of Different CE Strategies}
\label{sec:empirical baseline}
To systematically investigate performance differences among the three LLM-based CE strategies (RQ1), we first clarify our experimental setting: we focus on CE tasks and exclude cross-file repository context. This setting is rational because it isolates core code editing capabilities from irrelevant repository-level complexities, enabling fair horizontal comparison of the strategies' inherent code editing performance. Since the selected representative techniques for each strategy are originally designed for repository-level tasks, we adapt them to our scenario by retaining their core logic while removing repository-specific components.

\begin{itemize}[leftmargin=*]
    \item \textbf{CodePlan (One-step)} \cite{Bairi2024codeplan}: CodePlan takes the edit instructions and source code as input and integrates them with repository context and dependency analysis to directly generate fully modified code in a single pass. We preserve its intra-file context extraction (class structure sketching for target code blocks) and prompt construction mechanism, while removing repository-level components like incremental dependency graph construction and cross-file change propagation. 
    \item \textbf{AutoCodeRover (Search-replace)} \cite{Zhang2024autocoderover}: AutoCodeRover combines program structure search and fault localization to identify target code snippets, generates modification diffs, and replaces the corresponding regions in the original codebase. We preserve its AST-based intra-file code search, stratified context retrieval logic, and patch generation with retry-loop. We remove repository-level components including cross-file API invocation and multi-file fault localization.
    \item \textbf{FuseApply (Plan-apply)} \cite{Liang2025fuseapplybench}: FuseApply uses a large LLM to generate modified code drafts with placeholders, and a fine-tuned smaller LLM executes for the applying stage to fill in the placeholders with original code. This method natively takes the target single file and edit instruction as input, requiring no additional adjustments for our experimental scenario.
\end{itemize}

\subsubsection{Feature Analysis}
\label{sec:empirical feature analysis}
To investigate the features' influence on editing performance (RQ2), we employ a two-stage pipeline: LLM-based feature extraction to obtain candidate features from CE tasks and correlation-based validation to identify the significant influence of features.

\noindent{\textbf{LLM-based Feature Extraction.}}
We leverage LLMs to automatically extract features from our \benchmark by processing each CE task individually, capturing both surface-level statistics (e.g., lines of code) and semantic information (e.g., cross-function dependencies).
The extraction is guided by structured prompts that simulate the decision-making of human experts.
Specifically, the prompt comprises the following three components and is included in our release package~\cite{ourpackage}: 
\begin{itemize}[leftmargin=*]
    \item \textit{CE Strategy Introduction}: describes how each of the three CE strategies works, along with their advantages and disadvantages, helping the LLM understand the differences among strategies; 
    \item \textit{Task Assignment}: specifies the goal of identifying key features that influence strategy selection and explaining why they matter; 
    \item \textit{Input/Output Placeholder}: defines the input format (editing task description and code to be edited) and output structure (JSON format) to enable systematic feature collection.
\end{itemize}

The extraction process yielded 992 raw features. Considering that LLM-generated outputs often produce features with different literal names but similar semantics across tasks, we conduct semantic merging to unify overlapping and semantically identical features (e.g., ``total lines of code'' and ``file length in lines'').  Specifically, we adopt GPT-3.5 to identify semantically equivalent features. The LLM iteratively examines each group of extracted features, compares them with the existing feature set, and merges duplicates to remove items with identical meanings.
Despite the structural appearance of partial extracted features, the acquisition of these features hinges on the LLM's inherent semantic understanding. For instance, classifying task type (Bug Fix/Feature Add/Refactor) requires understanding the intent of the edit rather than simple syntactic pattern matching. 
Through this refinement process, we obtained the top ten most frequent features for subsequent analysis.

\noindent{\textbf{Correlation Validation.}} While the ten extracted features are frequently identified by LLMs as relevant to CE tasks, their actual influence on CE strategy selection remains unclear. To validate which features genuinely affect strategy choice, we conduct a correlation analysis to quantify the strength of association between each feature and strategy performance. 

First, we reformulate strategy selection as a binary classification task: for each CE task, the optimal CE strategy is labeled as 1, while the other two strategies are labeled as 0. This formulation enables us to independently evaluate each feature's association with each strategy.
Then, given the heterogeneous nature of the extracted features, we categorize them into numerical features (e.g., total lines of code and average modified lines per function) and categorical features (e.g., modification type). 
For numerical features, we employ Spearman's rank correlation coefficient~\cite{Spearman1904Proof} for correlation validation, which measures the monotonic relationship between feature values and CE strategy labels. We sort the numerical feature values in descending order and assign a unique rank to each value. Similarly, we apply this process to binary strategy labels. The Spearman coefficient $p$ is then computed based on these two sets of ranks.
For categorical features, we use the chi-square test~\cite{McHugh2013Chi}. Note that it evaluates whether feature categories and strategy outcomes are statistically independent, rather than individual strategies. Let $k$ denote the number of categories of the categorical feature,
we first construct a $k\times2$ contingency table where rows represent the feature and columns represent the binary strategy labels. The chi-square statistic $\chi^2$ is then calculated by comparing the observed frequencies in the table with the expected frequencies under the assumption of independence between the feature and the strategy. We set the significance level at $\alpha=0.05$, for $(k=1,\alpha=0.05,val=3.841)$, $\chi^2\geq val$ indicates a stronger association between the feature categories and the strategy.
Both methods follow established practices in multivariate analysis~\cite{Justin2016Multivariate,Shih2017Chi}.
Features with significant correlations with at least one strategy are retained as core explanatory variables for subsequent analysis.

\subsection{Evaluation Metrics}
\label{sec:empirical metirc}
We adopt the following metrics to evaluate both the correctness of editing tasks and token efficiency.

\begin{itemize}[leftmargin=*]
    \item \textbf{Pass@1:} It measures the proportion of tasks where CE strategies can successfully complete the task on the first attempt. We execute the edited code against test cases to assess the correctness of the edited code. Pass@1 is adopted as the primary metric since single-generation correctness aligns with real-world usage and ensures fair comparison across all methods.
    
    
    \item \textbf{Output Token:} It denotes the number of output tokens generated by the LLM-based editing strategies for each task. This metric serves as a proxy for token efficiency, i.e., lower output token counts indicate more cost-effective editing strategies. 
    
    \item \textbf{IsGT:} It is a binary indicator of whether a strategy achieves ground truth status on a given task. A strategy is considered ground truth if it satisfies two conditions: (1) it correctly completes the task, and (2) among all strategies that successfully complete the task, it generates the fewest output tokens. The IsGT identifies the unique optimal solution for each task.
    \end{itemize}

For feature analysis, we employ the following correlation coefficient metrics:
\begin{itemize}[leftmargin=*]
    \item \textbf{$P$-value:} It quantifies the statistical significance of the correlation between numerical features and strategies. We first compute the absolute values of Spearman's correlation coefficients between the numerical feature and each of the three strategies, then take the maximum value as the $P$-value. $P$-value exceeding 0.5 indicates a statistically significant correlation between the feature and strategy performance.
    \item \textbf{Cramer's $V$:} It measures the strength of association between multiclass features and strategy. We conduct a Chi-square test to assess the independence between the multiclass feature and strategy labels, then compute Cramer's $V$ based on the resulting Chi-square statistic and sample size. Cramer's $V\geq0.2$ indicates a non-negligible correlation between the feature and strategy selection, directly demonstrating that the feature exerts a meaningful impact on strategy selection.
\end{itemize}

\subsection{\benchmark Construction}
\label{sec: empirical data collection}

Existing code editing datasets often consist of simple tasks or lack an executable environment. Meanwhile, FuseApplyBench, a code editing dataset, primarily focuses on assessing error types during the code edit application process. 
To address this gap, we construct \benchmark by curating data from SWE-bench. This benchmark originates from real open-source projects and covers diverse practical code editing scenarios, with verified samples strictly checked manually.
Notably, SWE-bench includes a subset of highly challenging issues that cannot be resolved by mainstream code editing strategies. To enable the model to concentrate on code editing instead of file positioning, we filtered instances where groundtruth modifications are restricted to a single file and collected 500 CE tasks for our empirical study. In Section~\ref{sec:evaluation dataset}, we reconstruct a dataset \benchmarkV that includes both single-file and multi-file CE tasks for comprehensive evaluation.

We adopted the full content of the modified file as the original source code, and the SWE-bench problem statement as the editing requirement, thereby constructing structured code editing tasks. 

In addition to building the dataset, we also automatically labeled each task with its optimal editing strategy. Specifically, we evaluated three representative strategies on every instance. The raw outputs of these strategies were converted into patches and validated using the official SWE-bench evaluation framework, which ensures a fair and consistent measure of correctness. Alongside correctness, we recorded the Output Token of each strategy, to reflect their token efficiency. We designated the one with the lowest Output Token as the optimal strategy label.

Finally, we obtained \benchmark, which contains a total of 500 CE tasks. Among these tasks, 43 are labeled with one-step as the optimal strategy, 184 with search-replace as the optimal strategy, and 95 with plan-apply as the optimal strategy; the remaining 178 tasks are not labeled with an optimal strategy, as all three selected strategies failed to complete these tasks correctly.

\section{Empirical Result}

\begin{figure}[!t]
\centering

\begin{minipage}[c]{0.69\linewidth}  
\centering
\captionof{table}{\textbf{Overall Performance of Three Code Editing Strategies.} Pass@1 indicates the success rate of all editing tasks; Output Token Avg. measures the average output tokens of all
CE tasks; and \# of IsGT counts tasks achieving an optimal correctness-efficiency balance.}
\label{tab:method_performance}
 \resizebox{\textwidth}{!}{%
\begin{tabular}{
  p{2.0cm}<{\centering}  
  p{2.2cm}<{\centering}  
  p{1.0cm}<{\centering}  
  p{2.2cm}<{\centering}  
  p{0.6cm}<{\centering}  
} 
\toprule
\textbf{Approach} & \textbf{Strategy Type} & \textbf{Pass@1} & \textbf{Output Token Avg.} & \textbf{IsGT} \\ 
\midrule
CodePlan & One-step & 12.6 & 3,539 & 43 \\
AutoCodeRover & Search-replace & \textbf{48.2} & 1,268 & \textbf{184} \\
FuseApply & Plan-apply & 22.4 & \textbf{941} & 95 \\
\bottomrule
\end{tabular}
}

\end{minipage}
\hfill  
\begin{minipage}[c]{0.28\linewidth}
\centering
\includegraphics[width=\linewidth]{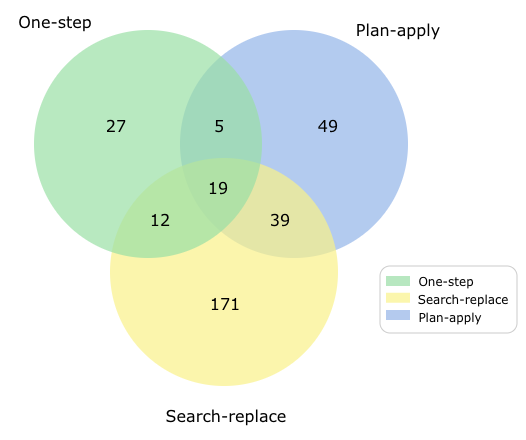}
\vspace{-7mm}  
\caption{Distribution of Successfully Solved Tasks Across Different Strategies}
\label{venn}
\end{minipage}
\vspace{-10pt}
\end{figure}

\begin{figure}[ht]
    \centering
    \includegraphics[width=1\linewidth]{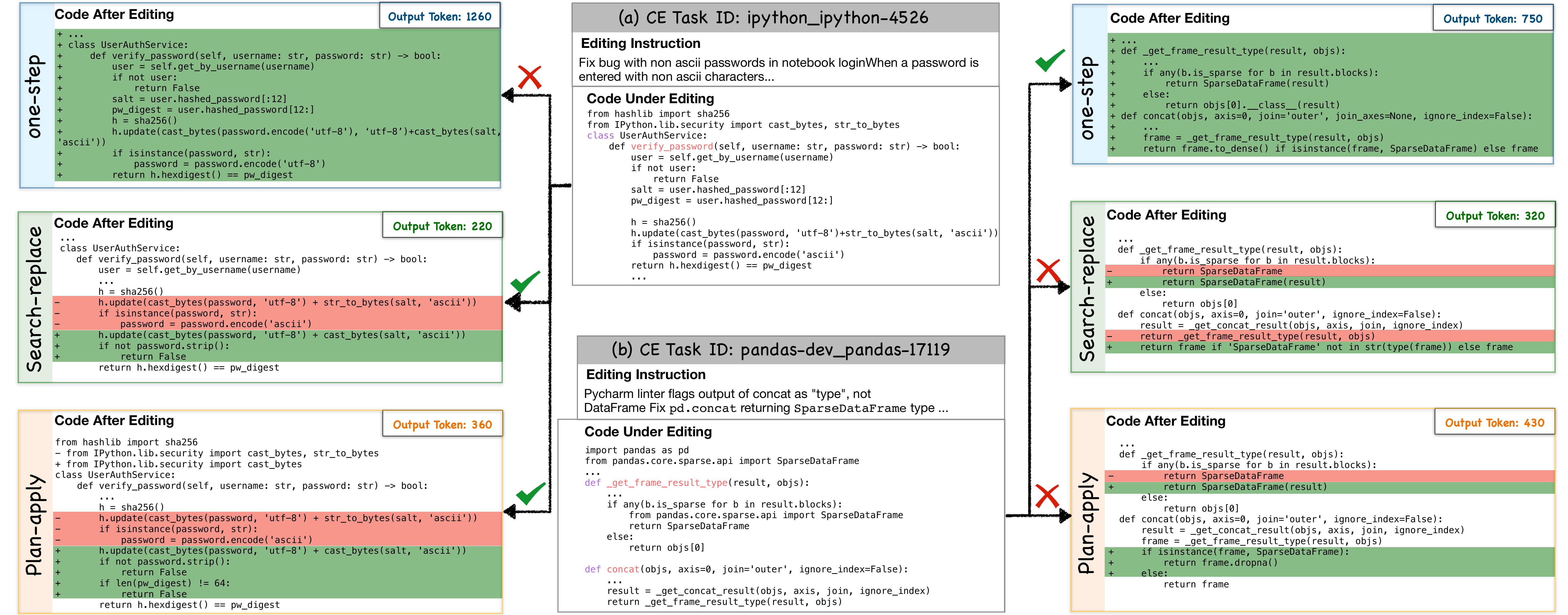}
    \vspace{-2mm}
     \caption{ Cases of Different Strategy Performance on  Specialized CE Tasks}
    \label{fig:case1}

\end{figure}

\subsection{Results of RQ1}

Tab.~\ref{tab:method_performance} presents the overall performance of three types of CE strategies across all tasks in \benchmark. We evaluate strategies using Pass@1 and Output Token. 
In Tab.~\ref{tab:method_performance}, AutoCodeRover (i.e., Search-replace) achieves the highest Pass@1 of 48.2\%, substantially outperforming FuseApply (i.e., plan-apply) and CodePlan (i.e., one-step).
This establishes search-replace as the most reliable strategy for completing CE tasks correctly. 
On the other hand, plan-apply demonstrates the best token efficiency with 941 average tokens, 25.8\% fewer than search-replace (1,268) and 73.4\% fewer than one-step (3,539). 
This indicates that plan-apply produces more concise edits. 
Search-replace achieves a balanced performance between correctness and efficiency for targeted edits. Fig.~\ref{fig:case1}-(a) illustrates this with an example: for a localized modification in a 120-line file, search-replace outputs the modified snippet (220 tokens) and completes the task correctly, while plan-apply, despite using placeholders, still consumes fewer tokens but fails more frequently.

The IsGT column in Tab.~\ref{tab:method_performance} and the successful task distribution in Fig.~\ref{venn} reveal critical insights into strategy complementarity. 
The overlapping regions in Fig.~\ref{venn} show that multiple strategies can correctly solve the same tasks. More importantly, the non-overlapping regions demonstrate that each strategy has exclusive successes, i.e., the tasks solvable only by that specific approach. Despite the poor overall Pass@1 (12.6\%) of the one-step strategy, it uniquely solves 17 tasks and serves as the ground truth for 43 tasks, revealing its benefits in certain scenarios.
Fig.~\ref{fig:case1}-(b)  illustrates such an example. For fixing $pd.concat$, which returned \texttt{SparseDataFrame} type, causing Linter errors, only the one-step strategy succeeded by instantiating sparse frames and adding dense conversion logic. Search-replace fixed Linter errors but introduced invalid type checks, while plan-apply corrected instantiation but misused \texttt{dropna()}, both failing to resolve the core issue. This confirms that no universal strategy exists, with each suited to specific task types.

\begin{tcolorbox}[colback=gray!10!white, colframe=gray!50!black, boxrule=0.5pt, before skip=3pt, after skip=3pt,left=5pt,        
  right=5pt,       
  top=5pt,         
  bottom=5pt ]
\textbf{\faLightbulbO~Findings of RQ1}: Our empirical results reveal two critical insights: (1) An effective CE strategy must prioritize task correctness while minimizing token cost; (2) No single CE strategy is universally optimal, and each strategy exhibits complementary strengths for different CE tasks.
\end{tcolorbox}

\subsection{Result of RQ2}

\begin{table}[!t]
\centering
\caption{Top Ten Frequent Features of Code Editing Tasks}  
\label{tab:feature_table}                  
\renewcommand{\arraystretch}{1.15}               
\resizebox{\linewidth}{!}{              \vspace{-10pt}
\begin{tabular}{l|l|l|c}
\toprule
\textbf{Feature} & \textbf{Type} & \textbf{Description} & \textbf{Frequency (\%)} \\
\midrule
Total Lines of Code          & Numerical          & Total lines of the target code file under edit& 69.46 \\
Modification Type            & Multiclass & Core edit type(Add/Delete/Modify) & 59.91 \\
Avg. Modified Lines per Function & Numerical      & Average lines modified per target function  & 15.38 \\
Modification Scope           & Numerical          & Number of functions that are modified & 12.35 \\
Dependency Relationship      & Numerical          & Number of functions that depend on modified functions & 6.53 \\
Task Complexity              & Numerical          & Modification volume (Added + Deleted + Modified Lines) & 6.53 \\
Modification Location        & Multiclass & Single function or not (Single/Multi) & 5.83 \\
Local Modification           & Multiclass          & Max continuous modified lines in file & 4.90 \\
Code Complexity     & Numerical          & Cyclomatic complexity of modified function & 4.90 \\
Task Type                   & Multiclass & Bug Fix/Feature Add/Refactor/Performance Optimization & 3.96 \\
\bottomrule
\end{tabular}
}
\vspace{-10pt}
\end{table}

\begin{table}[!t]
\centering
\caption{\textbf{Feature Correlation Coefficients.} Note: numerical features use Spearman's correlation coefficients, and multiclass features use Chi-square statistics.}
\vspace{-10pt}
\label{tab:feature_correlation}
\renewcommand{\arraystretch}{1.05}
\resizebox{1.0\linewidth}{!}{%
\begin{tabular}{l|l|c|c|c|c|c} 
\toprule
\textbf{Type} & \textbf{Feature} & \multicolumn{3}{c|}{\textbf{Correlation Coefficient}} & \textbf{P-value} & \textbf{Cramer's V} \\
\cmidrule(lr){3-5}
& & \textbf{One-step} & \textbf{Search-replace} & \textbf{Plan-apply} & & \\
\midrule
\multirow{7}{*}{Numerical} 
& \cellcolor{lb} Total Lines of Code & \cellcolor{lb}-0.65 & \cellcolor{lb}0.42 & \cellcolor{lb}0.38 & \cellcolor{lb}0.65 & \cellcolor{lb}- \\
\cmidrule{2-7} 
& \cellcolor{lb}Modification Scope & \cellcolor{lb}-0.60 & \cellcolor{lb}0.25 & \cellcolor{lb}0.36 & \cellcolor{lb}0.60 &\cellcolor{lb} - \\
\cmidrule{2-7}
& Avg. Modified Lines per Function & 0.48 & -0.42 & 0.45 & 0.48 & - \\
\cmidrule{2-7}
& Task Complexity & -0.23 & 0.18 & 0.09 & 0.23 & - \\
\cmidrule{2-7}
& Dependency Relationship & -0.22 & -0.15 & 0.09 & 0.22 & - \\
\cmidrule{2-7}
& Local Modification & 0.09 & 0.25 & -0.12 & 0.25 & - \\
\cmidrule{2-7}
& Code Complexity & 0.18 & -0.09 & -0.18 & 0.18 & - \\
\midrule
\multirow{3}{*}{Multiclass}
& \cellcolor{lb}Modification Type & \multicolumn{3}{c|}{\cellcolor{lb}32.64 (k=2, $\alpha=0.05$, val=5.99)} & \cellcolor{lb}- & \cellcolor{lb}0.32 \\
\cmidrule{2-7}
& \cellcolor{lb}Task Type & \multicolumn{3}{c|}{\cellcolor{lb}18.00 (k=3, $\alpha=0.05$, val=7.82)} & \cellcolor{lb}- &\cellcolor{lb} 0.24 \\
\cmidrule{2-7}
& Modification Location & \multicolumn{3}{c|}{3.08 (k=1, $\alpha=0.05$, val=3.84)} & - & 0.09 \\
\bottomrule
\end{tabular}
}
\vspace{-10pt}
\end{table}

As illustrated in RQ1, the performance of CE strategies is affected by task-specific scenarios. Hence, we further investigate which task-specific features account for this difference in performance.
Tab.~\ref{tab:feature_table} shows the top ten frequent features that LLMs considered for CE tasks.
Tab.~\ref{tab:feature_table} presents the ten most frequent features identified across CE tasks. \emph{Total lines of code} (69.46\%) and \emph{modification type} (59.91\%) emerge as the most prevalent features, appearing in over half of all tasks. Seven additional features show frequencies exceeding 5\%, while the majority of features appear in fewer than 5\% of tasks. 
Notably, high frequency does not necessarily indicate a strong impact on strategy selection, it merely reflects how often a feature appears across the dataset.

To assess actual impact, we perform correlation analysis on the ten most frequent features. As shown in Tab.~\ref{tab:feature_correlation}, we separate the ten features into two types for correlation validation. For numerical features (seven of the ten frequent features), we compute Spearman correlation coefficients and use the $p$-value to measure significance; for categorical features, we conduct Chi-square tests and measure Cramer's V value to assess significance. 
The validation reveals the following four features with significant influence on strategy effectiveness:
\begin{itemize}[leftmargin=*]
\item \textbf{Total Lines of Code} shows a general high correlation with strategy performance, with a $p$-value equal to 0.65. Additionally, strong negative correlation with one-step ($\rho$=-0.65) and positive correlation with search-replace ($\rho$=0.42) and plan-apply ($\rho$=0.38). This indicates that one-step performs better on smaller files, while search-replace and plan-apply are more suitable for larger codebases.
\item \textbf{Modification Scope} is also significant to strategy performance as $p$-value$= 0.6 > 0.5$. It exhibits a strong negative correlation with one-step ($\rho$=-0.60) and a positive correlation with plan-apply ($\rho$=0.36). Tasks requiring modifications across multiple functions favor plan-apply, while one-step is more effective for single-function edits.

\item \textbf{Avg. Modified Lines per Function} shows moderate correlation with all three strategies, with positive correlation for one-step ($\rho$=0.48) and plan-apply ($\rho$=0.45), but negative for search-replace ($\rho$=-0.42). This suggests search-replace excels when modifications are smaller per function.
\item \textbf{Modification Type} demonstrates moderate to strong association with strategy selection ($\chi^2 = 32.64 >5.99$, Cramer's V=0.32, p<0.05). Different modification types (Add/Delete/Modify) favor different strategies, indicating that the nature of code changes significantly influences optimal strategy choice.
\item \textbf{Task Type} shows moderate association with strategy effectiveness ($\chi^2 = 18.0 >7.82$, Cramer's V=0.24). Different task categories (Bug Fix/Feature Add/Refactor/Performance Optimization) exhibit varying strategy preferences, suggesting that the purpose of code editing affects strategy suitability.
\end{itemize}
Note that Task Types are not explicitly provided in the original dataset. Instead, we leverage an LLM to analyze the task type of each instance and conduct automatic clustering on all samples, and four typical task types are summarized accordingly: Bug Fix, Feature Add, Refactor and Performance Optimization. We further conduct manual examination to verify the classification quality and confirm that these four categories can cover more than 90\% of all tasks in our dataset.

The remaining features show weak correlations and no significant associations, indicating limited impact on strategy selection. 

\begin{tcolorbox}[colback=gray!10!white, colframe=gray!50!black, boxrule=0.5pt, before skip=3pt, after skip=3pt,left=5pt,        
  right=5pt,       
  top=5pt,         
  bottom=5pt ]
\textbf{\faLightbulbO~Findings of RQ2}: Strategy effectiveness is significantly influenced by four task-specific features: \emph{total lines of code}, \emph{modification scope}, \emph{modification type}, and \emph{task type}. This demonstrates that optimal CE strategy selection must be guided by task features rather than fixed approaches.
\end{tcolorbox}

\section{Methodology of \tool}
\begin{figure*}[h]
    \centering
    \includegraphics[width=1\linewidth]{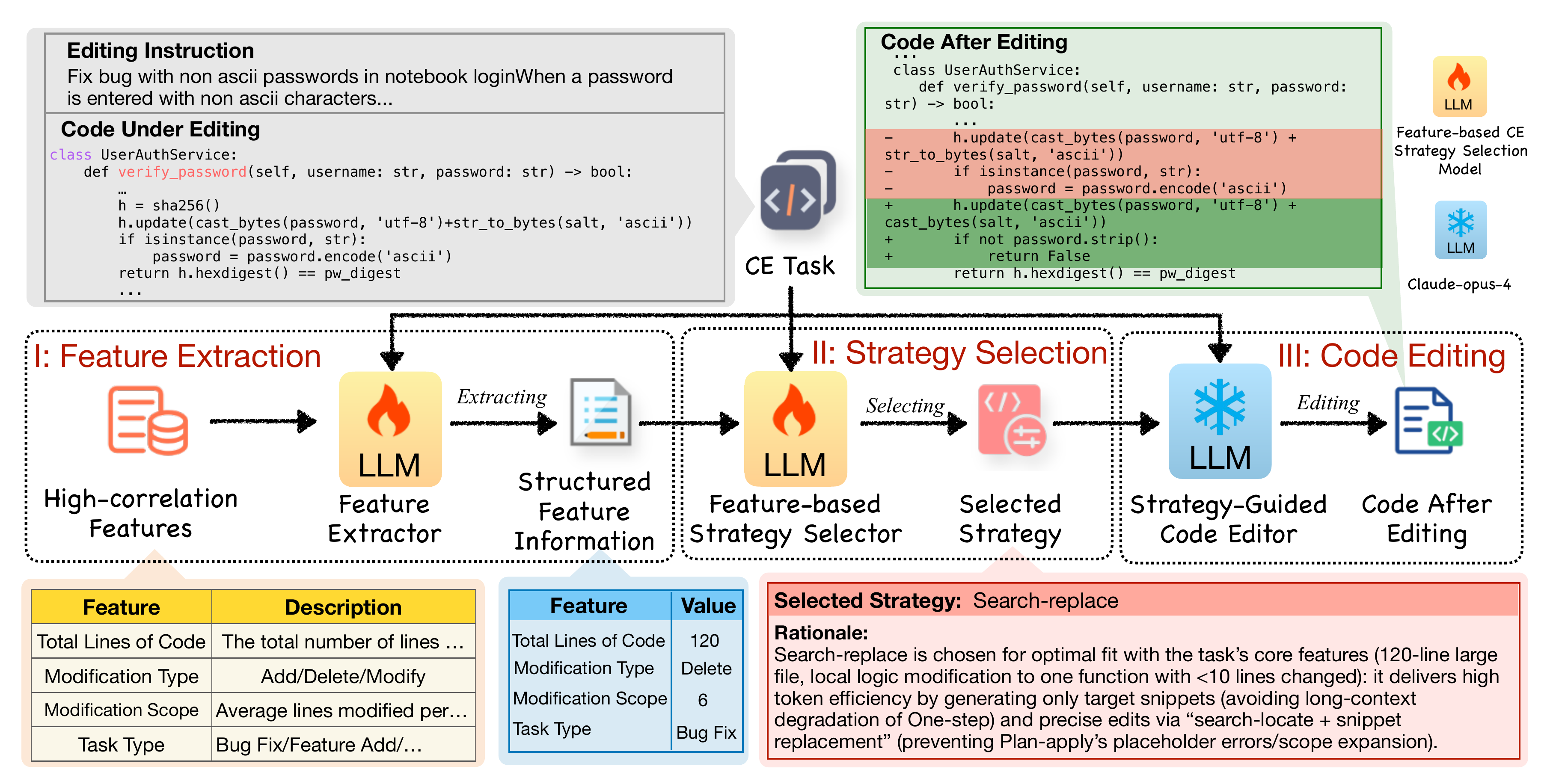}
    \vspace{-20pt}
    \caption{The overall framework of \tool and an illustrative case.}
    \label{fig:framework}
    \vspace{-10pt}
\end{figure*}

Motivated by our empirical findings that no single CE strategy is universally optimal and strategy effectiveness is significantly influenced by four task-specific features, we propose \tool, an adaptive framework that automatically selects optimal CE strategy from one-step, search-replace and plan-apply, based on task-specific features.
Building on these insights, we propose \tool, an adaptive framework that first automatically recommends the optimal strategy, along with interpretable explanations, for a given CE task, and
then employs LLMs to edit code. 
As illustrated in Fig.~\ref{fig:framework}, \tool operates through three phases: feature extraction (Phase I), strategy selection (Phase II), and code editing (Phase III). The core of \tool is a fine-tuned strategy selection model that learns feature-strategy mappings from our empirical study.

\subsection{Fine-tuned CE Strategy Selection Model}

Our empirical study reveals that optimal strategy selection depends on complex task-specific feature interactions. While general-purpose LLMs possess broad reasoning capabilities, they lack specialized knowledge to map feature patterns to strategy effectiveness. We therefore fine-tune Llama3.1-8B~\cite{grattafiori2024llama3herdmodels} using LoRA~\cite{Hu2021Lora}, which enables efficient adaptation while preserving general capabilities and offers better cost-efficiency than closed-source models.

The fine-tuning dataset integrates data from \benchmark and our findings of empirical study. Each training instance contains:
\begin{itemize}[leftmargin=*]
\item \textbf{Editing instruction:} Natural language description of the required code modification
\item \textbf{Code under editing:} The original source file requiring modification
\item \textbf{Feature information:} Quantified values of the four high-correlation features identified in RQ2 (i.e., \emph{total lines of code}, \emph{modification scope}, \emph{modification type}, and \emph{task type})
\item \textbf{Optimal strategy:} The strategy achieving task correctness and minimum output token cost, which is labeled in our \benchmark construction process.
\item \textbf{Rationale:} A structured natural language explanation that justifies the optimal strategy from feature perspectives, and elaborates the inherent deficiencies and inappropriateness of the other two CE strategies.

\end{itemize}
The construction of the rationale follows a fixed process. Given the extracted code features, source code, task, and optimal strategy for each instance, we instruct the LLM to generate explanatory text strictly from the feature perspective. Specifically, the LLM is tasked with analyzing the advantages and rationality of selecting the optimal strategy under the current feature conditions, and further illustrating the respective shortcomings and inapplicability of the other two candidate strategies. Subsequently, we perform manual inspection and minor revisions to ensure the reasonableness and factual consistency of all rationales. Figure~\ref{fig:framework} shows one of these examples.

This construction ensures the model learns from both empirical evidence and expert analysis, enabling informed and interpretable strategy selections.
We employ LoRA to fine-tune Llama3.1-8B on \benchmark with an 8:2 train-test split. The fine-tuning uses a learning rate of $ 1\times 10^{-5}$ and a batch size of 8, and the model is trained for 10 epochs to ensure sufficient exposure to strategy-selection examples while avoiding overfitting. Additionally, the LoRA rank is set to 32.

\subsection{Feature Extraction}

Given a CE task, \tool first extracts the high-correlation feature information using our fine-tuned CE strategy selection model in Phase I.
Specifically, the feature extraction phase includes three primary inputs: (1) \textbf{editing instruction} formulated in natural language; (2) \textbf{code under editing}, referring to the source code file requiring editing; and (3) \textbf{high-correlation CE features}, which are formal definitions and operational descriptions of the high-correlation features to be extracted identified through empirical study (RQ2).

With the input, the fine-tuned strategy selection model as the feature extractor performs automated analysis of the given CE tasks, and generates structured feature information. 
Specifically, the fine-tuned strategy selection model is guided by a feature-extraction instruction, which explicitly specifies the analysis procedure and quantification methodology for each feature.
Finally, the feature extractor processes the instructions and produces structured feature information containing quantified measurements across multiple dimensions relevant to strategy effectiveness. Fig.~\ref{fig:framework} also shows an example CE task focused on fixing the non-ASCII password bug in notebook login, where the original code's ASCII encoding for password processing fails to handle non-ASCII inputs.
With the given CE task (i.e., editing instruction and code under editing), the feature-based CE strategy selection model extracts the following high-correlation features information: $\{Total\ Lines\ of\ Code: 120;\ Modification\ Type: Delete;\ Modification\ Scope: 6;\ Task\ Type: Bug\ Fix\}$ in Phase I.

\subsection{Strategy Selection}
Phase II constitutes the core decision-making component of \tool, wherein the fine-tuned strategy selection model processes the extracted feature representations in conjunction with the CE task context to generate optimal strategy selections. 
This phase operationalizes the empirical knowledge encoded during fine-tuning process, applying learned feature-strategy mappings to novel CE tasks.

The strategy selection model receives a structured input comprising: (1) \textbf{editing instruction}, (2) \textbf{code under editing}, and (3) \textbf{structured feature information}, which is extracted in Phase I.
The input is formatted according to the schema employed during fine-tuning, ensuring distributional consistency between training and inference phases. 
This consistency is critical for enabling the model to effectively leverage the learned associations between feature patterns and strategy effectiveness.

With the input, we employ the fine-tuned model to perform strategy selection through the following multi-step reasoning process:
\textbf{(1) Feature pattern recognition:} Identifies feature combinations correlating with strategy effectiveness based on training instances. \textbf{(2) Contextual integration:} Examines feature interactions and holistic task context rather than individual values in isolation.
\textbf{(3) Strategy determination:} Selects the strategy predicted from one-step, search-replace, and plan-apply to maximize editing success and minimize output tokens.
\textbf{(4) Rationale generation:} Produces natural language explanation connecting features to strategy advantages.

For the case in Fig.~\ref{fig:framework}, the strategy selector recommends search-replace as the optimal strategy and generates the following rationale:
``\textit{Search-replace is chosen for optimal fit with the task's core features (120-line large file, local logic modification to one function with less than 10 lines changed): it delivers high token efficiency by generating only target snippets (avoiding long-context degradation of One-step), and avoids extraneous edits via `search-locate + snippet replacement'; (preventing Plan-apply's placeholder errors/scope expansion).}''
This rationale demonstrates the model's capacity to synthesize multiple features into a coherent decision framework.

\subsection{Code Editing}

In Phase III, \tool directs the LLM (Claude-Opus-4.1~\cite{Anthropic2025ClaudeOpus4}) to execute code editing following the selected strategy. 
Phase III translates the abstract strategy recommendation into concrete code modifications through LLM-based execution. 
\tool directs the LLM (i.e., claude-opus-4~\cite{Anthropic2025ClaudeOpus4}) to perform code editing in accordance with the strategy selected in Phase II, adhering to one of the three empirically-validated CE methodologies (i.e., one-step, search-replace, plan-apply) established in our empirical study.

In contrast to conventional CE approaches that employ static, strategy-agnostic prompting, \tool constructs an adaptive prompt that explicitly instructs the LLM to adhere to the selected strategy. The prompt integrates multiple information sources to provide comprehensive execution guidance, including (1) editing instruction, (2) code under editing, (3) selected strategy, (4) strategy rationale, and (5) strategy-specific execution protocol.
The editing instruction and code under editing are original data in the constructed benchmark; the selected strategy and rationale are generated by Phase I and II; and the execution protocols are as follows:
\begin{itemize}[leftmargin=*]
\item \textbf{One-step protocol:} It directs LLMs to generate the complete modified code file in a single response while preserving the overall code structure, context, and dependencies. 
\item \textbf{Search-replace protocol:} It directs the LLM to first identify the precise code segments requiring modification, then provide exact search-replace pairs for each target location.
\item \textbf{Plan-apply protocol:} It directs the LLM to decompose the editing task into two sequential phases: first generating a modification plan with placeholders, then populating the placeholders with actual code to produce the final result. 
\end{itemize}

Our \tool finally returns the code after editing, along with the selected strategy and the corresponding rationale, enabling developers to make informed decisions about accepting or refining the automated code editing.
For example in Fig.~\ref{fig:framework}, with the selected strategy (i.e., search-replace), \tool identifies the target code snippet:\texttt{
h.update(cast\_bytes(password,'utf-8')+str\_to\_bytes(salt, 'ascii'))}
And generates the cryptographically-secure replacement:

\begin{center}
\vspace{-0.2cm}
\begin{minipage}{1\textwidth} 
\begin{lstlisting}[language=Python]
Search:     
       h.update(cast_bytes(password,'utf-8')+str_to_bytes(salt, 'ascii'))
       if isinstance(password, str):
           password=password.encode('ascii')
Replace:
       h.update(cast_bytes(password,'utf-8')+cast_bytes(salt,'ascii'))
       if not password.strip():
           return False
\end{lstlisting}
\vspace{-0.2cm}
\end{minipage}
\end{center}
This precise search-replace approach ensures surgical modification of only the intended code segment, preventing unintended alterations to surrounding code and maintaining the functional integrity of the remainder of the function.

\section{Evaluation Setup}


\subsection{Research Questions}
To evaluate the effectiveness of our \tool, we investigate the following research questions:
\begin{itemize}[leftmargin=*]
    \item \textbf{RQ3: How effective is \tool in performing code editing tasks compared to baseline approaches, and how accurate is our strategy selection?} To address RQ3, we compare \tool against three categories of baselines: (1) fixed single-strategy approaches, (2) classifier-based strategy selection, and (3) instruction-prompting methods. We evaluate \tool from two perspectives, i.e., \textit{task performance}, measured by task correctness (Pass@1) and token efficiency, and \textit{strategy selection quality}, measured by the accuracy of recommended CE strategies and the corresponding rationales.
    \item \textbf{RQ4: What are the contributions of individual components and features in \tool?} To investigate RQ4, we conduct ablation studies by separately removing components and features, then evaluating their contributions to the overall framework's performance.
    \item \textbf{RQ5: How well does \tool generalize across different LLMs for code editing?} We conduct a comprehensive evaluation by replacing LLMs in Phase III with a diverse set of state-of-the-art LLMs, encompassing both closed-source models (Claude-Opus-4.1~\cite{Anthropic2025ClaudeOpus4}, GPT-4o~\cite{zhang2025gpt4ogoldstandardscalable}, Gemini-2.5-Pro~\cite{GoogleDeepMind2025Gemini2.5Pro}) and open-source models (Qwen3-235B-A22B~\cite{qwen3technicalreport}, DeepSeek-R1~\cite{deepseekai2025deepseekr1incentivizingreasoningcapability}, Llama-3.1-405B~\cite{grattafiori2024llama3herdmodels}). We compare \tool against AutoCodeRover, a state-of-the-art approach that employs a fixed search-replace strategy, measuring both code editing correctness (Pass@1) and token efficiency (Output Token count).

\end{itemize}

\subsection{Evaluation Dataset}
\label{sec:evaluation dataset}
To ensure rigorous and unbiased evaluation of \tool, we construct a new evaluation dataset named \benchmarkV, derived from SWE-bench and SWE-bench-verified~\cite{jimenez2024swebench}. For single-file tasks, we follow the same construction methodology as \benchmark (see Sec.~\ref{sec: empirical data collection}). To further comprehensively assess the practical effectiveness of our \tool, we extend the dataset to support multi-file editing scenarios. For multi-file instances, we extract all source files and corresponding file paths covered by the ground-truth patch as the input source code, and take the problem statement as the editing instruction. We curate 650 CE tasks, including 429 single-file and 221 multi-file instances. Among these tasks, 39 have one-step as the optimal strategy, 152 have search-replace as the optimal strategy, 126 have plan-apply as the optimal strategy, and 333 tasks are marked with N/A. For tasks marked as N/A, we exclude them from the Select@1 evaluation, as the optimal strategy cannot be reliably determined among the given candidate strategies. These instances are still included for computing Pass@1 and average output tokens.  We verify that \benchmarkV is strictly disjoint from the fine-tuning dataset to prevent data leakage, ensuring that our evaluation results reflect the framework's generalization capability in unseen code-editing scenarios rather than memorization of the fine-tuning data.

\subsection{Baselines}\label{sec:baseline} 
To evaluate the performance of our \tool, we design two baseline approaches: a fixed single-editing strategy and a selective editing strategy.
For the fixed single-editing strategy, we select three representative CE strategies in Sec.~\ref{sec:empirical baseline}, i.e., CodePlan~\cite{Bairi2024codeplan}, AutoCodeRover~\cite{Zhang2024autocoderover}, and FuseApply~\cite{Liang2025fuseapplybench}.
For selective editing strategy, we propose the following strategy-selection approaches as baselines. 
\begin{itemize}[leftmargin=*]
    \item \textbf{LLM-based Instruction-driven Selector (LLM-IS)}: With the taxonomy of the three CE strategies (i.e., one-step, search-replace, and plan-apply), we first instruct LLMs to select an appropriate CE strategy based on the task description and source code. Then, we adopt the selected strategy for completing the CE task. Here, we also employ Claude-Opus-4.1 as the selector. Different from our \tool, LLM-IS is not provided with explicitly extracted features during strategy selection, and it relies solely on the raw source code and task to make its decision.
    \item \textbf{CodeBERT-based Strategy Classifier (CodeBERT-SC)}: We utilized \benchmark to train a classifier on CodeBERT~\cite{Guo2020graphcodebert} to map CE task features with one of the three CE strategies. Subsequently, the classified strategy is adopted to instruct LLMs to complete the CE task.
\end{itemize}

To investigate the contribution of the framework design, we utilize variants for an ablation study:
\begin{itemize}[leftmargin=*]
    \item \textbf{\tool$_{w/o\ FE}$:} It is a variant of \tool by removing the feature extraction in Phase I, where the fine-tuning CE strategy selection model selects a strategy according to edit instruction and raw code under editing, without given structured high-correlation feature information.
    \item \textbf{\tool$_{w/o\ SS}$:} It is designed for evaluating the contribution of strategy selection in Phase II. We consider two alternatives: \ding{172} Removing the strategy selection phase, and directly instructing the extracted structured feature information into phase III for code editing.
    We adopt search-replace as the editing strategy, and do not provide any rationale for guidance;
    \ding{173} Replace the fine-tuned strategy selection model with a closed-source model for strategy selection, to contrast the performance of a fine-tuned small model versus a large closed-source model.
    \item \textbf{\tool$_{w/o\ Features}$:} We also remove individual features from our process to assess their impact, including: \ding{172} removing total lines of code from the structured feature information, \ding{173} removing modification type, \ding{174} removing average modification length per function, and \ding{175} removing modification scope.
\end{itemize}

\subsection{Evaluation Metrics}
We use \textbf{Pass@1} and \textbf{Output Token} to evaluate the correctness and effectiveness of the CE task, respectively.
The description and calculation methods for these two are the same as those in Sec.~\ref{sec:empirical metirc}.

Additionally, we also evaluate the accuracy of strategy selection and corresponding explanations.
For strategy selection accuracy, we compute \textbf{Select@1}, which is the proportion of tasks where the strategy selection model correctly identifies the optimal editing strategy on the first attempt. 
For the selection rationale, two authors independently evaluate the \textbf{accuracy} of the generated rationale for each CE task. 
They judge whether or not the generated rationales are correct. 
We then compute Cohen's Kappa~\cite{Kappa1996richard} to evaluate the inter-rater agreement. 
For the rationales that the
two authors disagree, they have to discuss and come to a consensus.
Based on consensus annotations, we evaluate the quality of the rationale generated for each CE task.

\subsection{Experimental Environment}
 We conduct all experiments on a Linux machine with Ubuntu 20.04.4LTS. It is equipped with an Intel(R)Xeon(R)Platinum 8,358P CPU of 2.60 GHZ with 128 cores and 2 TB memory. 
 For all baselines, the experiments adopt Claude-opus-4-1-20250805~\cite{Anthropic2025ClaudeOpus4} in Phase III as the inference model.
 For Phase I and II, we use Llama3.1-
 8B~\cite{grattafiori2024llama3herdmodels} to fine-tune the CE strategy selection model.

\section{Evaluation Result}
\subsection{Overall Effectiveness and Strategy Selection Quality (RQ3)}

\begin{table}[!t]
\centering
\caption{Overall Performance of \tool in code editing tasks;
\footnotesize{Values are reported as \textbf{Overall (Single-file/Multi-file)}, $\uparrow$ / $\downarrow$ denote the relative improvement of \tool over each baseline.}}
\label{tab:overall_performance}
\renewcommand{\arraystretch}{0.8}
\vspace{-5pt}
\resizebox{1.0\linewidth}{!}{
\small
\begin{tabularx}{1\linewidth}{
  >{\hsize=1.00\hsize\centering\arraybackslash}X
  >{\hsize=0.9\hsize\centering\arraybackslash}X
  >{\hsize=1.00\hsize\centering\arraybackslash}X
  >{\hsize=1.15\hsize\centering\arraybackslash}X
  >{\hsize=1.00\hsize\centering\arraybackslash}X
  >{\hsize=0.95\hsize\centering\arraybackslash}X
} 
\toprule
\textbf{Approach} & \textbf{Strategy Type} & \textbf{Pass@1} & \textbf{Output Token Avg.} & \textbf{Select@1} & \textbf{Rationale Acc.}\\ 
\midrule
CodePlan & Fixed (One-step) & 11.8 (14.8/5.9) (\textcolor{red}{$\uparrow$ 294.9\%}) & 4253 (3722/5285) (\textcolor{red}{$\downarrow$ 68.2\%})& - & - \\ \midrule
AutoCodeRover & Fixed (Search-replace) & 38.3 (46.1/23.1) (\textcolor{red}{$\uparrow$ 21.7\%}) & 1861 (1575/2417) (\textcolor{red}{$\downarrow$ 27.3\%})  & -  & -\\ \midrule
FuseApply & Fixed (Plan-apply) & 24.3 (29.2/14.9) (\textcolor{red}{$\uparrow$ 91.8\%}) & \textbf{1253 (1102/1545)} (\textcolor{red}{$\uparrow$ 7.9\%}) & - & -\\\midrule
LLM-IS & Adaptive & 26.6 (31.2/17.6) (\textcolor{red}{$\uparrow$ 75.2\%}) & 1921 (1762/2231) (\textcolor{red}{$\downarrow$ 29.6\%}) & 34.1 (32.9/36.4) (\textcolor{red}{$\uparrow$ 112.9\%}) & 25.9 (24.7/28.3) (\textcolor{red}{$\downarrow$ 68.2\%})\\ \midrule
CodeBERT-SC & Adaptive & 30.7 (35.8/20.8) (\textcolor{red}{$\uparrow$ 51.8\%}) & 1617 (1436/1967) (\textcolor{red}{$\downarrow$ 16.4\%})& 43.9 (44.3/43.2) (\textcolor{red}{$\uparrow$ 65.4\%}) & -\\ \midrule
\cellcolor{lb}\tool & \cellcolor{lb} Adaptive & \cellcolor{lb}\textbf{46.6(53.1/33.9)} & \cellcolor{lb}1352 (1160/1724) & \cellcolor{lb}\textbf{72.6(73.1/71.6)} & \cellcolor{lb}\textbf{69.9(70.4/68.9)}\\ 
\bottomrule
\end{tabularx}
}
\vspace{-5pt} 
\end{table}

Table~\ref{tab:overall_performance} presents the comparative results of \tool against all baseline approaches on \benchmarkV. We analyze the results from two perspectives: code editing performance (i.e., task correctness and token efficiency), and strategy selection quality.

\noindent\textbf{Code Editing Performance.}
\tool achieves an overall Pass@1 of 46.6\%, with 53.1\% on single-file tasks and 33.9\% on multi-file tasks, substantially outperforming all baselines. Compared to fixed-strategy approaches, \tool shows improvements of 294.9\% over CodePlan, 21.7\% over AutoCodeRover, and 91.8\% over FuseApply. Among adaptive selection methods, \tool outperforms LLM-IS by 75.2\% and CodeBERT-SC by 51.8\%. These results demonstrate that our fine-tuned strategy selection model effectively identifies appropriate strategies for diverse CE tasks.
The performance gap between fixed strategies reveals the importance of adaptive selection. AutoCodeRover (38.3\%) performs best among fixed approaches, but still falls 8.3 percentage points short of \tool. 

Regarding token efficiency, \tool consumes an average of 1,352 output tokens overall, with 1,160 tokens on single-file tasks and 1,724 tokens on multi-file tasks, achieving near-optimal efficiency. While FuseApply uses fewer tokens (1,102 on single-file tasks and 1,545 on multi-file tasks, with an overall average of 1,253), its overall Pass@1 is only 24.3\% (29.2\% on single-file tasks), representing a poor correctness-efficiency trade-off. Compared to other adaptive methods, \tool reduces token consumption by 29.6\% relative to LLM-IS (1,921 overall) and 16.4\% relative to CodeBERT-SC (1,617 overall), while maintaining superior correctness. This demonstrates that \tool's strategy selection not only improves task success rates but also promotes efficient token usage.

\noindent\textbf{Strategy Selection Quality.}
\tool achieves an overall Select@1 of 72.6\%, with 73.1\% on single-file tasks and 71.6\% on multi-file tasks, significantly outperforming LLM-IS (34.1\%) by 112.9\% and CodeBERT-SC (43.9\%) by 65.4\%. This substantial improvement demonstrates the effectiveness of our fine-tuned model compared with instruction-prompting and CodeBERT-based classification methods.
The low selection accuracy of LLM-IS indicates that general-purpose LLMs, without training and explicit structured features, struggle to make optimal strategy decisions based solely on instructions.
CodeBERT-SC performs better by leveraging task features, but remains limited by its shallow semantic understanding and its inability to capture complex task-strategy relationships. 
In contrast, \tool's fine-tuned model learns nuanced associations between task features and strategy appropriateness from a diverse fine-tuning dataset, resulting in more accurate selections.

Unlike most baselines, our fine-tuning strategy selection model also provides rationales for its selections.
To evaluate rationale quality, two authors independently assessed tasks in CE tasks from \benchmarkV, achieving substantial inter-rater agreement (Cohen's $\kappa$=0.807). Consensus annotations showed that 69.9\% of \tool's rationales were accurate, significantly outperforming LLM-IS (25.9\%). The improvement demonstrates that fine-tuning enables more accurate task-strategy associations than instruction-prompting LLMs. 
The overall selection accuracy (72.6\%) combined with high rationale accuracy (69.9\%) indicates that \tool not only makes correct strategy decisions but also provides interpretable justifications.

\begin{tcolorbox}[colback=gray!10!white, colframe=gray!50!black, boxrule=0.5pt, before skip=3pt, after skip=3pt,left=5pt,        
  right=5pt,       
  top=5pt,         
  bottom=5pt ]
\textbf{\faPencil~Answer to RQ3}: \tool significantly outperforms all fixed-strategy and adaptive baseline approaches in both editing correctness and token efficiency. Its fine-tuned strategy selector achieves high selection accuracy with well-justified rationales, validating the effectiveness of our approach for adaptive code editing strategy selection.
\end{tcolorbox}

\subsection{Ablation Study (RQ4)}

\begin{table}[!t]
\centering
\caption{Ablation Study of \tool; \footnotesize{Values are reported as \textbf{Overall (Single-file/Multi-file)}, $\uparrow$ / $\downarrow$ denote the relative improvement of each variant compared with \tool.} }
\label{tab:ablation_ace}
\renewcommand{\arraystretch}{1.1}
\vspace{-5pt}
\small
\begin{tabular}{
  p{0.20\linewidth}
  |c|
  c
  c
} 
\toprule
\multicolumn{1}{c|}{\textbf{Approach}} & \textbf{Alternative} & \textbf{Pass@1} & \textbf{Output Token Avg.} \\
\midrule

\cellcolor{lb}\centering\textbf{\tool} & \cellcolor{lb} -& \cellcolor{lb}\textbf{46.6 (53.1/33.9)} & \cellcolor{lb}\textbf{1352 (1160/1724)} \\
\midrule

\centering \tool$_{w/o\ FE}$ & - & 23.0 (26.3/16.6) (\textcolor{red}{$\downarrow$ 50.6\%}) & 3032 (2565/3939) (\textcolor{red}{$\uparrow$ 124.3\%}) \\
\midrule

\multirow{4}{*}{\centering \quad\quad \tool $_{w/o\ Feature}$} 
& \ding{172} & 34.7 (39.5/25.4) (\textcolor{red}{$\downarrow$ 27.0\%}) & 2525 (2149/3255) (\textcolor{red}{$\uparrow$ 86.8\%}) \\ \cline{2-4}
& \ding{173} & 35.3 (41.2/23.8) (\textcolor{red}{$\downarrow$ 24.2\%}) & 1743 (1477/2259) (\textcolor{red}{$\uparrow$ 28.9\%})\\ \cline{2-4}
& \ding{174} & 43.2 (49.4/31.2) (\textcolor{red}{$\downarrow$ 7.3\%}) & 1381 (1192/1748) (\textcolor{red}{$\uparrow$ 2.1\%}) \\ \cline{2-4}
& \ding{175} & 44.6 (51.0/32.2) (\textcolor{red}{$\downarrow$ 4.3\%})& 1454 (1247/1856) (\textcolor{red}{$\uparrow$ 7.5\%})\\
\midrule

\multirow{2}{*}{\centering \quad\quad \tool$_{w/o\ SS}$} 
& \ding{172} & 39.0 (45.2/27.0) (\textcolor{red}{$\downarrow$ 16.3\%}) & 1801 (1534/2319) (\textcolor{red}{$\uparrow$ 33.2\%})\\ \cline{2-4}
& \ding{173} & 35.2 (40.7/24.5) (\textcolor{red}{$\downarrow$ 24.5\%}) & 1895 (1610/2448) (\textcolor{red}{$\uparrow$ 40.2\%})\\
\bottomrule
\end{tabular}
\vspace{-10pt}
\end{table}

To understand the contributions of individual components and features in \tool, we conduct systematic ablation studies. Table~\ref{tab:ablation_ace} presents the results of our ablation study.

\noindent\textbf{Impact of Feature Extraction.}
From Tab.~\ref{tab:ablation_ace}, we find that removing feature extraction (\tool$_{w/o\ FE}$) causes the most severe degradation, with the overall Pass@1 decreasing by 50.6\% and the average output tokens increasing by 124.3\%. Correspondingly, token consumption rises from 1160 to 2565 (single-file) and from 1724 to 3939 (multi-file).
This demonstrates that explicit feature extraction is the most critical component.
Without structured feature information, the model cannot effectively select appropriate strategies for both single-file and multi-file editing tasks, leading to lower correctness and significantly higher token consumption.

\noindent\textbf{Impact of Individual Features.}
We also systematically remove individual features to assess their relative importance. Removing \emph{Total Lines of Code} (\ding{172}) and \emph{Modification Type} (\ding{173}) cause the largest impacts, with Pass@1 decreasing by 27.0\% and 24.2\%, respectively. These features are most critical for strategy selection as they directly determine whether holistic or targeted strategies are appropriate. In contrast, removing \emph{Average Modification Length} (\ding{174}) and \emph{Modification Scope} (\ding{175}) result in smaller impacts (7.3\% and 4.3\% decreases), indicating they provide complementary but less essential information. This indicates a feature hierarchy where code magnitude and modification type are primary factors, with modification extent features serving as secondary refinements.

\noindent\textbf{Impact of Strategy Selection.}
The strategy selection module is indispensable to the overall framework. When strategy selection is removed and replaced with search-replace strategy without rationale guidance (\tool$_{w/o\ SS}$ \ding{172}), the overall Pass@1 drops by 16.3\%. This reveals that even with extracted features available, adopting a fixed search-replace strategy without explicit adaptive strategy guidance is insufficient to guarantee editing performance.
Replacing the fine-tuned model with few-shot prompting (\tool$_{w/o\ SS}$\ding{173}) results in even greater degradation (24.5\% decrease), validating our choice to use a fine-tuned model that internalizes feature-strategy mappings from 322 empirical cases rather than relying on general-purpose prompting.

\begin{tcolorbox}[colback=gray!10!white, colframe=gray!50!black, boxrule=0.5pt, before skip=3pt, after skip=3pt, left=5pt,        
  right=5pt,       
  top=5pt,         
  bottom=5pt ]
\textbf{\faPencil~Answer to RQ4}: Ablation study confirms that: (1) feature extraction is the most critical component; (2) code magnitude and modification type are the most important features for guiding CE tasks; (3) the fine-tuned strategy selection model significantly outperforms few-shot prompting. These findings validate the essential design decisions in \tool.
\end{tcolorbox}

\subsection{Generalization Across Different LLMs (RQ5)}

\begin{table}[!t]
\centering
\caption{Performance Comparison Between Open-Source and Closed-Source Models; \footnotesize{Values are reported as \textbf{Overall (Single-file/Multi-file)}, $\uparrow$ / $\downarrow$ denote the relative improvement of \tool compared with AutoCodeRover.}}

\label{tab:model_comparison}
\small
\begin{tabular}{
p{2.0cm}<{\centering}
|p{2.0cm}<{\centering}
|p{1.8cm}<{\centering}
|p{2.0cm}<{\centering}
|p{1.9cm}<{\centering}
|p{2.0cm}<{\centering}
}
\toprule
\textbf{Category} & \textbf{LLM} & \multicolumn{2}{c|}{\textbf{AutoCodeRover}} & \multicolumn{2}{c}{\textbf{\tool}} \\
\cline{3-6}  
& & \textbf{Pass@1} & \textbf{Token Avg.} & \textbf{Pass@1} & \textbf{Token Avg.} \\
\midrule
\multirow{3}{*}{\textbf{Closed-source}} 
& Claude-Opus-4.1 
& 38.3(46.1/23.1) 
& 1861(1575/2417) 
& 46.6(53.1/33.9) (\textcolor{red}{↑21.7\%}) 
& 1352(1160/1724) (\textcolor{red}{↓27.3\%})\\ \cline{2-6}

& GPT-4o 
& 28.8(34.7/17.4) 
& 1944(1645/2525) 
& 36.5(41.6/26.5) (\textcolor{red}{↑26.7\%}) 
& 1569(1347/2002) (\textcolor{red}{↓19.2\%}) \\ \cline{2-6}

& Gemini-2.5-Pro 
& 27.0(32.5/16.3) 
& 1886(1596/2450) 
& 31.7(36.2/23.1) (\textcolor{red}{↑17.4\%}) 
& 1536(1319/1960) (\textcolor{red}{↓18.6\%}) \\
\midrule
\multirow{3}{*}{\textbf{Open-source}} 
& DeepSeek-R1 
& 33.8(40.7/20.4) 
& 1954(1654/2539)
& 40.0(45.6/29.1) (\textcolor{red}{↑18.3\%}) 
& 1490(1279/1901) (\textcolor{red}{↓23.7\%}) \\ \cline{2-6}

& Qwen3-235B-A22B 
& 25.7(30.9/15.6) 
& 1984(1679/2577) 
& 30.6(34.9/22.3) (\textcolor{red}{↑19.1\%})
& 1538(1320/1962) (\textcolor{red}{↓22.4\%})\\ \cline{2-6}

& Llama-3.1-405B 
& 27.3(32.9/16.4) 
& 1918(1623/2491) 
& 35.0(39.9/25.5) (\textcolor{red}{↑28.2\%})
& 1522(1307/1942) (\textcolor{red}{↓20.6\%}) \\
\bottomrule
\end{tabular}
\vspace{-10pt}
\end{table}

To assess the generalizability of our \tool, we employ different LLMs for code editing in Phase III, including three closed-sourced LLMs and three open-sourced LLMs.
For evaluation, we compare the CE correctness (i.e., Pass@1) and effectiveness (i.e., Output Token) with the SOTA fixed strategy CE approach (i.e., AutoCodeRover), which uses a search-replace editing strategy.

Table~\ref{tab:model_comparison} illustrates that \tool consistently outperforms AutoCodeRover across all six LLMs in both correctness and cost efficiency. 
For correctness, \tool achieves Pass@1 improvements ranging from 17.4\% to 28.2\%, with an average improvement of 21.9\%. Notably, the improvements are consistent across both closed-source and open-source models, demonstrating that the adaptive strategy selection approach generalizes effectively regardless of the underlying LLM's architecture or training methodology.
For token efficiency, \tool reduces output token consumption by 18.6\% to 27.3\% compared to AutoCodeRover, with an average reduction of 22.0\%. This substantial reduction indicates that by selecting task-appropriate strategies, \tool generates more concise and targeted code modifications, avoiding the unnecessary token overhead associated with fixed-strategy approaches. 
The token efficiency gains are slightly differentiated between model categories: for closed-source models, \tool achieves an average 21.7\% reduction (27.3\% for Claude-Opus-4.1, 19.2\% for GPT-4o, 18.6\% for Gemini-2.5-Pro), compared to an average 21.9\% reduction for open-source models (23.7\% for DeepSeek-R1, 22.4\% for Qwen3-235B-A22B, 20.6\% for Llama-3.1-405B).
These results demonstrate that \tool's feature-based adaptive strategy selection provides consistent benefits across diverse LLMs with varying capabilities, confirming the framework's broad applicability and robustness. The improvements hold for both high-performing models (e.g., Claude-Opus-4.1) and relatively weaker models (e.g., Qwen3-235B-A22B), indicating the strategy selection mechanism effectively enhances CE performance independent of base model quality.

\begin{tcolorbox}[colback=gray!10!white, colframe=gray!50!black, boxrule=0.5pt, before skip=3pt, after skip=3pt,left=5pt,        
  right=5pt,       
  top=5pt,         
  bottom=5pt ]
\textbf{\faPencil~Answer to RQ5}: Our \tool demonstrates robust generalization across diverse LLMs. Across six LLMs (three closed-source, three open-source), \tool consistently outperforms the fixed-strategy baseline with 21.9\% average improvement in Pass@1 and 22.0\% reduction in token consumption, confirming the broad applicability of adaptive strategy selection.
\end{tcolorbox}

\section{Discussion}
\subsection{Latency and Token Consumption}

Ace employs a locally deployed, fine-tuned lightweight model (Llama 3.1-8B) for feature extraction and strategy selection (Phase I and Phase II), with an average latency of only ~600 milliseconds per task and no additional API token cost. In Phase III, the extra input overhead to guide the LLM in executing the selected strategy averages fewer than 600 tokens. In contrast, LLM-based adaptive selection baselines (LLM-IS) rely on external closed-source LLM API calls for decision-making, incurring 1,500–2,000 additional tokens per selection and ~2–3 seconds of extra network latency. The modest latency overhead of Ace’s strategy selection phase is fully offset by a 27.3\% reduction in output token consumption relative to AutoCodeRover during code editing, yielding superior end-to-end efficiency in real-world code editing tasks in terms of both latency and token cost.

\subsection{Threats to Internal Validity}

\noindent\textbf{Task Scope Limitation.}
To enable robust comparison of core editing capabilities, we removed repository-level context from original methods, focusing exclusively on file-level editing logic. This simplification may not fully reflect real-world coding scenarios, but it effectively isolates the core editing strategies from confounding repository-related factors, ensuring that observed performance differences stem directly from the strategies' inherent designs rather than external complexities.

\noindent\textbf{Token Efficiency Measurement.}
We measure token efficiency using output tokens only, excluding input and intermediate reasoning tokens. This design is justified by two considerations: (1) input tokens are uniform across all strategies in our experiments, and (2) intermediate reasoning tokens are implementation-specific and not part of the final editing output. Since our focus is evaluating the efficiency of code editing results rather than internal reasoning processes, this metric aligns with our research objectives of comparing output efficiency across strategies.

\noindent\textbf{Manual Annotation Bias.}
Two stages involve manual annotation: feature quantification from LLM-extracted factors and rationale quality assessment. Subjective differences in annotator interpretation may introduce bias. To mitigate this threat, we employ multiple measures: (1) two independent annotators with consensus resolution for disagreements, (2) Cohen's Kappa to quantify inter-rater reliability (0.80, indicating substantial agreement), and (3) systematic ablation experiments validating the impact of identified features on strategy performance.

\noindent\textbf{LLM Hallucination.}
Hallucination is a fundamental property of the underlying LLM rather than a behavior specific to any particular code editing strategy.
In this work, we use Pass@1 as an indirect indicator of hallucination risk, since any hallucinatory behavior will lead to test case failures.
The consistent superiority of Ace over all baselines in Pass@1 demonstrates that our feature-based adaptive strategy selection does not exacerbate hallucination compared to baselines.

\subsection{Threats to External Validity}

\textbf{Adaptability to Different Data.} Both feature extraction and evaluation were conducted on Python code from \benchmark, meaning the extracted features are closely aligned with Python’s syntactic structures and code organization. Other programming languages differ substantially in syntax and project structure. 
This indicates that \tool's performance may decline when applied to non-Python languages due to limited feature adaptability, potentially restricting cross-language generalization. Additionally, while \benchmark focuses on real-world software development tasks, the identified features may not capture key characteristics of other scenarios such as code completion~\cite{deng2025enhancing} or competitive programming~\cite{jain2025livecodebench}, which have distinct requirements. Nevertheless, our feature-based methodology remains applicable to arbitrary CE datasets and programming languages.

\section{Conclusion}
 In this paper, we address the ``one-size-fits-all'' limitation in CE tasks by conducting a systematic empirical study. We constructed \benchmark, including 500 real-world CE tasks, to compare three representative CE strategies (one-step, search-replace, plan-apply). Our findings confirm no single strategy is universally optimal, with four core CE task features: total lines of code, modification type, modification scope, and task type.
 Building on these insights, we proposed \tool, an adaptive code editing framework that integrates feature extraction, strategy selection, and code editing. Extensive evaluations on \benchmark-verified show \tool outperforms state-of-the-art baselines, improving the performance of various LLMs for CE tasks.
 \tool provides an effective solution for real-world CE tasks. In future work, we plan to further improve methods for enhancing code editing efficiency.

\section*{Data Availability}
Our prompt, code, dataset, and experimental results are available at \cite{ourpackage}.

\begin{acks}
This research/project is supported by the National Natural Science Foundation of China (No.62302437, No.92582107), Yongjiang Talent Program (No.2023A-402-G), and Ant Group.
\end{acks}

\bibliographystyle{ACM-Reference-Format}
\bibliography{software}
\end{document}